\documentclass[10pt,letterpaper]{article}
\usepackage[top=0.85in,left=2.75in,footskip=0.75in]{geometry}
\usepackage{changepage}
\usepackage[utf8x]{inputenc}
\usepackage{textcomp,marvosym}
\usepackage{fixltx2e}
\usepackage{amsmath,amssymb}
\usepackage{cite}
\usepackage{nameref,hyperref}
\usepackage[right]{lineno}
\usepackage{microtype}
\DisableLigatures[f]{encoding = *, family = * }
\raggedright
\usepackage[aboveskip=1pt,labelfont=bf,labelsep=period,justification=raggedright,singlelinecheck=off]{caption}

\makeatletter
\renewcommand{\@biblabel}[1]{\quad#1.}
\makeatother

\date{}

\usepackage{lastpage,fancyhdr,graphicx}
\usepackage{epstopdf}
\fancyheadoffset[L]{2.25in}
\fancyfootoffset[L]{2.25in}
\begin{document}
\vspace*{0.2in}
\begin{flushleft}
{\Large
\textbf\newline{Transcriptional and translational regulation in Arc protein network of {\textit {\textbf  Escherichia coli}}'s stress response} 
}
\newline
\\
Sutapa Mukherji,
\\
\bigskip
 Department of Protein Chemistry and Technology, 
Central Food Technological Research Institute, 
Mysore-570 020, India
\\
\bigskip

* sutapa@cftri.res.in

\end{flushleft}
\section*{Abstract}
In the recent past, there has been a lot of effort  in understanding sRNA mediated regulation of gene 
expression and how this mode of regulation differs from transcriptional regulation. 
 In {\textit {\textbf  Escherichia coli}}, in the presence of oxidative stress, 
  the synthesis of the sigma factor,  $\sigma^s$, 
is regulated  through   an interesting mechanism wherein sRNA mediated translational regulation 
is combined with transcriptional regulation. 
The key regulatory factors involved in transcriptional and translational regulation are  
ArcA and ArcB proteins and  a small regulatory RNA (sRNA), ArcZ,  respectively.  
The phosphorylated ArcA protein functions through a feedforward mechanism wherein it 
represses the transcription of   ArcZ  sRNA, a post-transcriptional activator for  
$\sigma^s$ and also  directly represses the transcription of $\sigma^s$. Through a feedback mechanisms, ArcZ sRNA 
destabilises ArcB mRNA and thus regulates the concentration of ArcB protein
which is a kinase that phosphorylates    ArcA.  The oxygen and energy availability is expected to influence the 
ArcA phosphorylation rate and,  as a consequence, in equilibrium, the system is likely  to be in 
either of the two states with high ArcB (low ArcZ) or low ArcB (high ArcZ)   concentration. 
In this work, we formulate a mathematical model that combines two distinctly different mechanisms 
of regulation at transcriptional and translational levels.   Kinetic modelling studies show that   the 
rate of destabilisation of ArcB mRNA by ArcZ sRNA  must be appropriately tuned for the cell 
to achieve the desired state. In particular, in case of a high phosphorylation rate,  
the  transition from a low ArcZ synthesis regime to a high ArcZ synthesis 
 regime with the increase in  sRNA-mRNA  interaction is similar to  the 
 threshold-linear response observed earlier. Further, it  is shown that the 
 sRNA mediated mRNA destabilisation might be, in particular, 
 beneficial in the low phosphorylation state  for 
 having the right   concentration levels of  ArcZ  and   ArcB.  
Stochastic  simulations carried out here   suggest that as the  ArcZ-ArcB binding affinity is increased,  
the probability distribution for the number of ArcZ molecules  becomes flatter indicating frequently occurring 
transcriptional bursts of varying strengths. Such a behavior might be  useful for switching of the cell from one state to 
to another.   

\section*{Introduction}
\label{sec:introduction}
 The role of  small regulatory RNAs  in the regulation of protein synthesis has been explored in various contexts. 
   So far,  approximately $100$ sRNAs have been identified in {\it Escherichia coli} .  
  In the recent past, there have been several evidences  that these sRNAs regulate  protein 
  synthesis by decreasing the  mRNA stability  \cite{masse,masse2} or repressing  the  translational activity \cite{aiba,wagner}.   
 In certain cases, there are reports of   increase in target protein concentrations  
 due to positive regulation by sRNAs \cite{vasudevan, pillai, ma,gadgil, papenfort}. sRNA inhibits  translation initiation
 by base-pairing with the mRNA  at the ribosome binding site.    In case of translational activation, it promotes ribosome binding by 
 exposing the binding site. In many cases,  sRNAs bind tightly to other molecules such as RNA chaperone Hfq which facilitate sRNA 
 binding with mRNA molecules, protect sRNA from degradations etc. 
Since  sRNA molecules are small (50-250 nucleotides long for bacteria) and 
 are not translated, it is believed that sRNA provides a  cost  effective and rapid way of regulation of gene expression. 
 Although transcriptional regulation by proteins has been under extensive investigation,  post transcriptional  regulation by sRNA is 
 relatively less understood. 
 It is expected that the non-catalytic nature of sRNA-mRNA interaction would be qualitatively different in nature from the 
 catalytic protein-DNA interaction. 
 This is also  evident from some of the recent investigations  \cite{hwa,hwa2}  that report 
  threshold-linear response with tunable threshold,  smoothening of gene expression  etc., in sRNA mediated gene regulation.  
 In addition,  quantitative differences in the time scale  of response in case of  protein-DNA, sRNA-mRNA and 
  protein-protein interactions  have been reported in \cite{shimoni}.

In {\it E. coli}, in the presence of oxidative stress, the synthesis of $\sigma$ factor, $\sigma^s$,  is  regulated  through an 
interesting mechanism that involves regulation both at  transcriptional and  translational levels. In general, 
the  sigma factors are some of the key stress response regulators of the bacterial cells since  they
play an important role in switching  various transcription initiation programs of the  cells. RNA polymerase (RNAP) 
holoenzymes (E$\sigma$), formed upon binding of the core RNA polymerase to a specific sigma factor,  are 
responsible for the transcription initiation. 
$\sigma$ factors   can be of 
different kinds such as $\sigma^{\rm 70},\ \sigma^s, \sigma^{\rm 32}$ etc. and  a
specific promoter recognition by $E\sigma$ depends on the particular sigma subunit 
 $E\sigma$ has. It is found  that 
$\sigma^s$  or the RpoS subunit  acts as  a  key regulator of   stress response  
of  {\it E. coli} \cite{hengge}.  As a consequence of a high cell density or environmental stress, 
rapidly multiplying bacterial cells enter into a  stationary 
phase. In the stationary phase,   $\sigma^s$ 
is overexpressed  compared to the housekeeping $\sigma$ factor $\sigma^{\rm 70}$ 
which plays a significant role in the  growth phase. 
Measurements of levels of  free or bound RNAPs and 
various sigma factors and subsequent modelling seem to indicate that various 
$\sigma$ factors compete for  binding to the core RNAP \cite{grigorova,klumpp}.  
The over-expression of  $\sigma^s$ in the stationary phase 
thus results in an increase in  the level of $E\sigma^s$ holoenzyme 
in comparison to other forms of $E\sigma$. 
Once bound to the RNAP, $\sigma^s$ regulates the expression of 
approximately $500$  genes 
responsible for bacterial survival under stress 
conditions.
 
In regulatory networks, certain specific  patterns of (sub)network are often found 
in abundance  in comparison to other  networks that could be 
constructed, in principle, by  joining various nodes with edges \cite{mangan-alon,rosenfeld, shen-orr,bose}.  
These frequently occurring patterns of (sub)networks  
are usually referred as network motifs.  Examples of these motifs include 
autoregulatory  loops, feedforward loops, feedback loops  etc. An autoregulatory loop  is the 
one in which  a gene product regulates the expression of its own gene. In 
case of positive (or negative)  autoregulation the gene product activates ( or represses) 
 the transcription of its own gene. Feedforward loops are the ones in which a transcription  factor  
 X regulates the transcription of Z directly and also in an indirect  way through the regulation of the 
 transcription of Y which in turn  regulates the transcription of Z.  Thus there is a direct as well as an   
 indirect path  of regulation (via regulation of Y) of Z  mediated by X. Similar feedforward loop 
 formed by sRNA mediated regulation has also been observed recently \cite{papenfort2}.
   The fact that these network 
motifs  occur more frequently than  other possible  designs of networks indicates that  
their presence   is beneficial for  regulation in comparison to other  networks \cite{alon}. 
 The $\sigma^s$ regulatory network described below involves two such nontrivially coupled 
 network motifs involving transcriptional and translational regulation.  
 
\begin{figure}[ht!]
  \centering
   \includegraphics[height=0.6\textwidth]{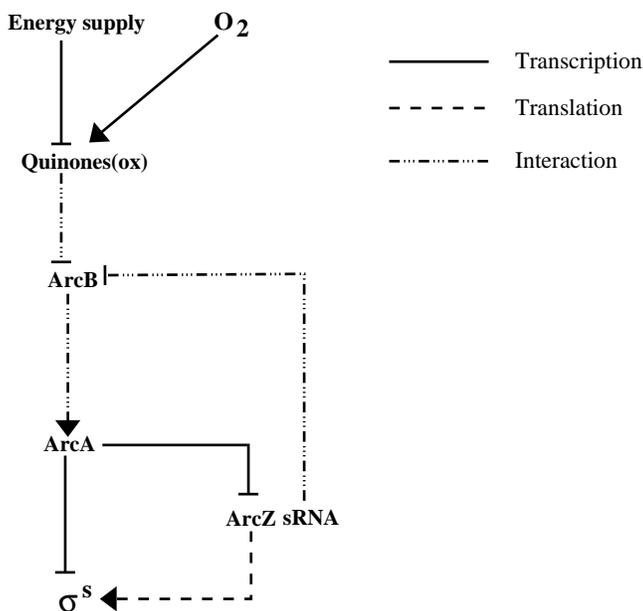}  
\caption{ArcA/ArcB/ArcZ regulation in $\sigma^s$ synthesis. 
 Lines ending with a bar represent repression. Arrowed lines represent activation. }
\label{fig:samplefig1}
\end{figure}

The cytosolic response regulator, ArcA, and the   sensor kinase, ArcB,  are known to play 
important roles in rapid response to oxygen availability to the cell \cite{georgellis,lin,lynch,peercy,kiley}.
A two component system formed of    ArcA and ArcB  has been  found to  regulate   
$\sigma^s$ synthesis in {\it E.coli} for stress under oxygen and energy supply \cite{mika}. 
Further, recent experiments suggest  that  in addition to ArcA and ArcB, there exists also 
 a third regulator,  a small regulatory RNA (sRNA) known as  ArcZ  \cite{gottesman}. 
  As shown in figure (\ref{fig:samplefig1}),  the network 
comprises of   coupled  feedforward and   feedback loops  involving  the three key regulators ArcA, ArcB and 
ArcZ. The sensor kinase, ArcB, upon autophosphorylation,  phosphorylates ArcA. The level of 
phosphorylated ArcA  depends on the oxygen availability  since oxidised  quinones  inhibit
  the autophosphorylase activity of ArcB. 
Further,  the phosphorylation of ArcA  depends on the energy supply since, apart from oxygen supply,  
 the  redox state of quinone   also depends on  the  energy available  to the system. 
The feedforward mechanism is due to the  phosphorylated ArcA which regulates $\sigma^s$ synthesis in 
two ways. It  represses  the transcription of 
$\sigma^s$ directly,   and   also  affects   the synthesis of $\sigma^s$  indirectly 
through  the transcriptional 
  repression of  ArcZ  which can activate   $\sigma^s$ translation \cite{gottesman}.  
In addition to this,  there is 
 a feed back loop  through which ArcZ down regulates  ArcB 
 by destabilising  ArcB mRNA.     Intuitively, 
the design of the network shows the possibility of two distinct  states.
 A state with a high level of ArcB  and, consequently, a low  level of ArcZ, 
 is  expected when the phosphorylation rate is high. 
 With a  low phosphorylation rate, we 
 expect a state with low ArcB and high ArcZ concentrations. 

 The main purpose of this study is to combine two distinct regulation mechanisms involving catalytic and 
 non-catalytic interactions to  explore the role of sRNA mediated regulation over  the 
 transcriptional  regulation  by phosphorylated ArcA. Kinetic modelling based analysis suggests that, 
among all the interactions,  the sRNA  mediated post transcriptional repression,  might be the most crucial  one 
 for reaching the desired equilibrium concentration of ArcZ or ArcB.  Further, we  show that the sRNA mediated 
 regulation might be, in particular,  necessary  in the low phosphorylation state.  This is due to the fact that 
 with the lowering of the phosphorylation rate,   the  ArcZ concentration saturates to 
 an upper limit dependent on its synthesis and   degradation rate.  
 Whether  the ArcZ concentration exceeds that of   ArcB protein  depends on  these rates as well as the  synthesis 
 and degradation rates of ArcB mRNA and   ArcB protein.  
 In such a scenario, the sRNA mediated ArcB destabilisation appears  as an additional mechanism  
 to further increase the ArcZ concentration. In the high phosphorylation state, 
 the sRNA-mRNA interaction might not be as crucial as mentioned above since in this case 
  ArcZ concentration can reduce to as small level as required by increasing the phosphorylation rate alone. 
 To study how the stochastic fluctuation in sRNA synthesis is influenced by its auto regulation, we have 
 performed stochastic simulations.  Average probability distributions of ArcZ molecules over a 
 fixed time interval  are  obtained for different ArcZ-ArcB binding affinities.  It appears that as the 
 sRNA-mRNA interaction increases, the probability distribution becomes flatter indicating large
 number of transcriptional bursts of varying strengths. This feature might have deeper implications 
 in the  context of switching of the cell from one  state to another \cite{madanbabu}.

 \section*{Methods}
 \label{sec:mathematical}
\begin{figure}[ht!]
  \centering
   \includegraphics[height=0.6\textwidth]{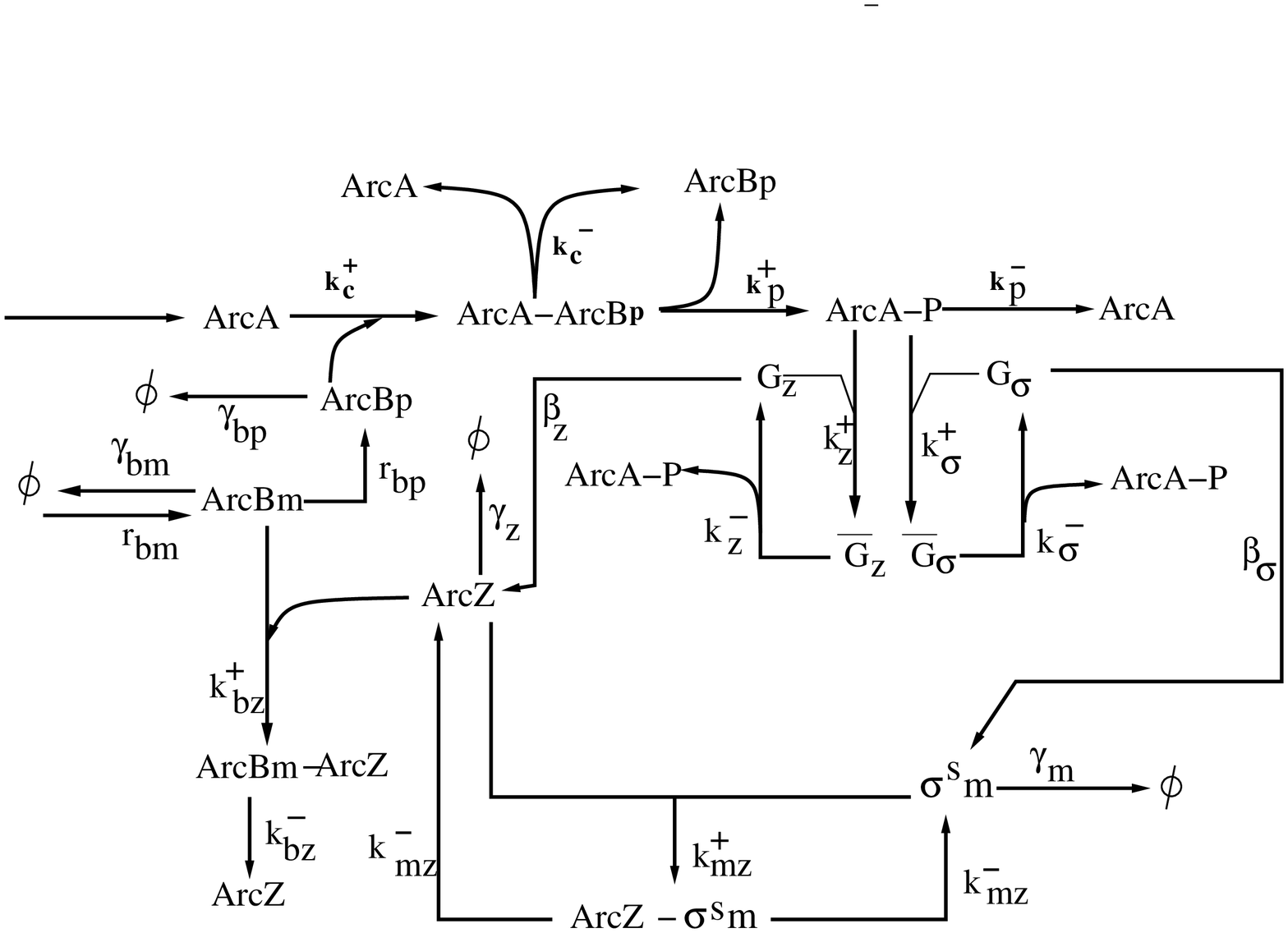}  
\caption{The network considered for  mathematical modelling.  A fixed concentration of  ArcA has been considered here. }
\label{fig:workingnetwork}
\end{figure}

 In the kinetic modelling scheme, we write the   time evolution of average concentrations of various proteins, mRNAs etc.  
 in terms of  differential equations. 
All the reactions along with the respective  rate parameters are presented in figure (\ref{fig:workingnetwork}).

The time evolutions of ArcB mRNA and ArcB protein  concentrations  can be described as  
\begin{eqnarray}
\frac{d[B_m]}{dt}=-\gamma_{bm} [B_m]+r_{bm}-k_{bz}^+ [B_m][Z] \label{diffeqnbm}\\
\frac{d[B_p]}{dt}=-k_c^+ [B_p][A]+k_c^-[AB_p]+k_p^+[AB_p]-\gamma_{bp} [B_p]+r_{bp} [B_m] \label{diffeqnbp}
\end{eqnarray}
where $[B_m]$, [Z] and $[B_p]$ represent  concentrations of ArcB mRNA, ArcZ sRNA and  ArcB protein, respectively. 
Here, $[A]$, $[AB_p]$ denote the concentrations of ArcA and ArcA, ArcB protein
 complexes, respectively. 
Further, $r_{bm}$ and $\gamma_{bm}$ denote the rate of supply and degradation, respectively, of ArcB mRNA; 
 $r_{bp}$ and $\gamma_{bp}$ represent the rate of synthesis and degradation, respectively,  
  of   ArcB protein. $k_{bz}^+$ denotes the rate of ArcB mRNA, ArcZ sRNA complex formation. 
  Equilibrium  concentrations of various  components  can be   obtained by solving the  nonlinear 
  equations obtained upon  equating various   time derivatives  to zero. 
 Using  equilibrium condition for the  phosphorylation process (details provided in  the  Supporting Information), 
we find from equations   (\ref{diffeqnbm}) and (\ref{diffeqnbp})
\begin{eqnarray}
&&[B_m]=\frac{r_{bm}}{k_{bz}^+[Z]+\gamma_{bm}} \ \ {\rm and} \label{steadybm}\\
&& [B_p]=\frac{r_{bp}}{\gamma_{bp}}[B_m].\label{steadybp}
\end{eqnarray}

Assuming the synthesis rate of ArcZ sRNA and $\sigma^s$mRNA from active states of the respective genes to be 
$\beta_z$ and $\beta_\sigma$, respectively, we have 
\begin{eqnarray}
\frac{d[Z]}{dt}=\beta_z \frac{G_z^{\rm tot}}{1+k_z [AP]}-k_{bz}^+ [B_m][Z]+k_{bz}^-[B_mZ]-k_{mz}^+ [\sigma^sm][Z]+
k_{mz}^- [\sigma^smZ]-\gamma_z [Z]\label{diffeqnz}\\
\frac{d[\sigma^sm]}{dt}=\beta_\sigma \frac{G_\sigma^{\rm tot}}{1+k_\sigma [AP]}+k_{mz}^- [\sigma^smZ]-
k_{mz}^+[\sigma^sm][Z]-\gamma_m [\sigma^sm]\label{diffeqnsigmam}\\
\frac{d[B_mZ]}{dt}=k_{bz}^+ [B_m] [Z]-k_{bz}^-[B_mZ]\label{diffeqnbz}\\
\frac{d[\sigma^smZ]}{dt}=k_{mz}^+ [\sigma^sm][Z]-k_{mz}^-[\sigma^sm Z], \label{diffeqnsigmamz}
\end{eqnarray}
where $k_\sigma= \frac{k_\sigma^+}{k_\sigma^-}$ and 
 $k_z=\frac{k_z^+}{k_z^-}$ (see  Supporting Information).
Further, $[\sigma^sm]$, $[B_mZ]$ and $[\sigma^smZ]$ represent concentrations of $\sigma^s$ mRNA, bound complexes of 
ArcB mRNA  and  ArcZ sRNA,  bound complexes of  $\sigma^s$ mRNA and ArcZ sRNA, respectively.   The first terms in (\ref{diffeqnz}) and 
(\ref{diffeqnsigmam}) represent the repression  activity of phosphorylated ArcA in  {\it arcZ} and $\sigma^s$ expression, respectively with 
 $G_z^{\rm tot}$ and $G_\sigma^{\rm tot}$ denoting the total number of the copies of {\it arcZ} and $\sigma^s$ gene, respectively.
Equilibrium  conditions on (\ref{diffeqnbz}), (\ref{diffeqnsigmamz}) and (\ref{diffeqnz}) together imply 
\begin{eqnarray}
\beta_z \frac{G_z^{\rm tot}}{1+k_z [AP]}-\gamma_z[Z]=0\label{solnz1}.
\end{eqnarray}

\section*{Results}
\subsection*{Steady state solution}
Using the equilibrium solution for $[AP]$ (see Supporting Information) and equations 
(\ref{steadybm}) and (\ref{steadybp})  for  $[B_p]$ and $[B_m]$, we have  from (\ref{solnz1})
\begin{eqnarray}
\gamma_z [Z]\left[1+\left( \frac{r_{bp}}{\gamma_{bp}}\right)\frac{k_{AP} k_z r_{bm} [A]}{k_{bz}^+[Z]+\gamma_{bm}}\right]=\beta_z G_z^{\rm tot}. \label{zsolution}
\end{eqnarray}
 Equation (\ref{zsolution}) can be solved 
for the equilibrium  ArcZ concentration. The solution for $[Z]$ is 
\begin{eqnarray}
[Z]=[Z^*]=\frac{1}{2 k^+_{bz}}\bigg\{-(\gamma_{bm}+K[A]-\frac{\beta_z}{\gamma_z} G_z^{\rm tot} k_{bz}^+)+ \big [(\gamma_{bm}+K[A]-
\frac{\beta_z}{\gamma_z} G_z^{\rm tot} k_{bz}^+)^2+\nonumber\\
4 \frac{\gamma_{bm}}{\gamma_z}\beta_z G_z^{\rm tot} k_{bz}^+\big ]^{1/2}\bigg\},
\label{steadyz}
\end{eqnarray}
where $K=k_z k_{AP} r_{bm} r_{bp}/\gamma_{bp}$.  Here, $k_{AP}=\frac{k_p k_c}{1+k_p^+/k_c^-}$ with $k_p=\frac{k_p^+}{k_p^-}$, $k_c=\frac{k_c^+}{k_c^-}$. 
Although the binding affinity between  ArcZ and  $\sigma^s$ transcripts 
does not influence  the equilibrium concentrations of ArcZ or ArcB, these concentrations 
 depend on  $k_{bz}^+$, the binding affinity of ArcB mRNA and ArcZ sRNA.  
 We show  in the next subsection that  the interaction between  ArcZ sRNA and  $\sigma^s$ mRNA, however,  
 influences  the time 
scale over which the system reaches  equilibrium.   

\subsection*{Numercal solutions for various concentrations}
\begin{figure}[!htb]
  \minipage{0.5\textwidth}
 \includegraphics[height=0.55\textwidth]{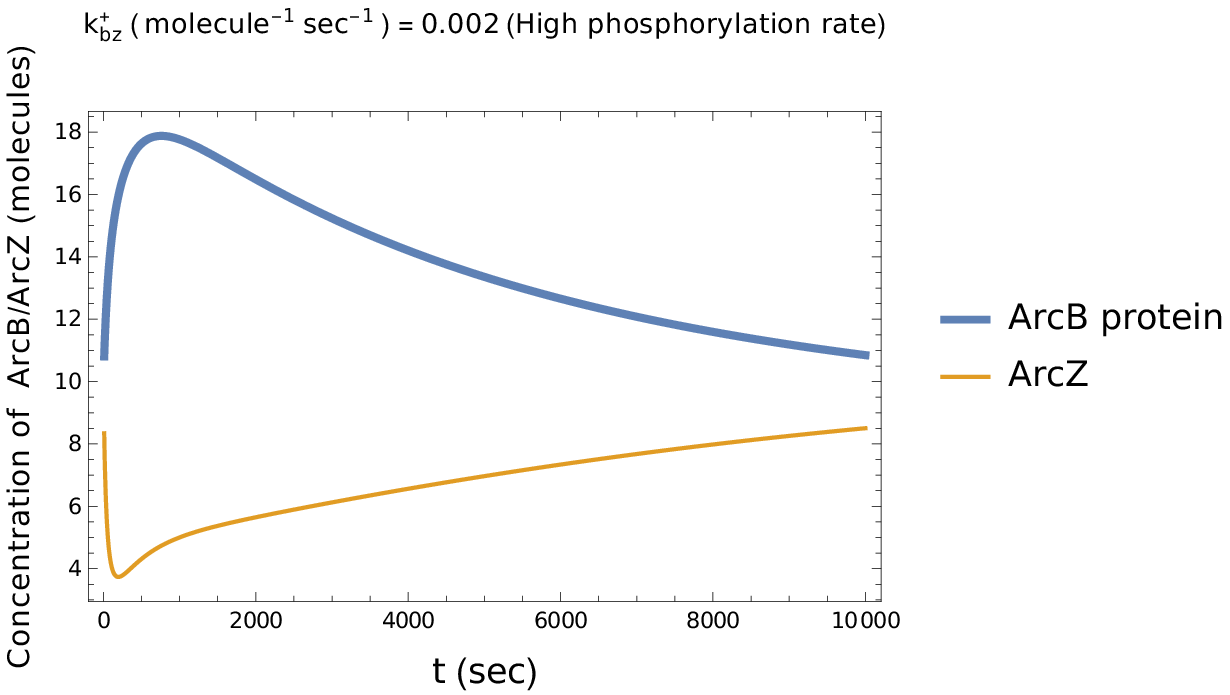}
 \endminipage
 \minipage{0.5\textwidth}
  \includegraphics[height=0.55\textwidth]{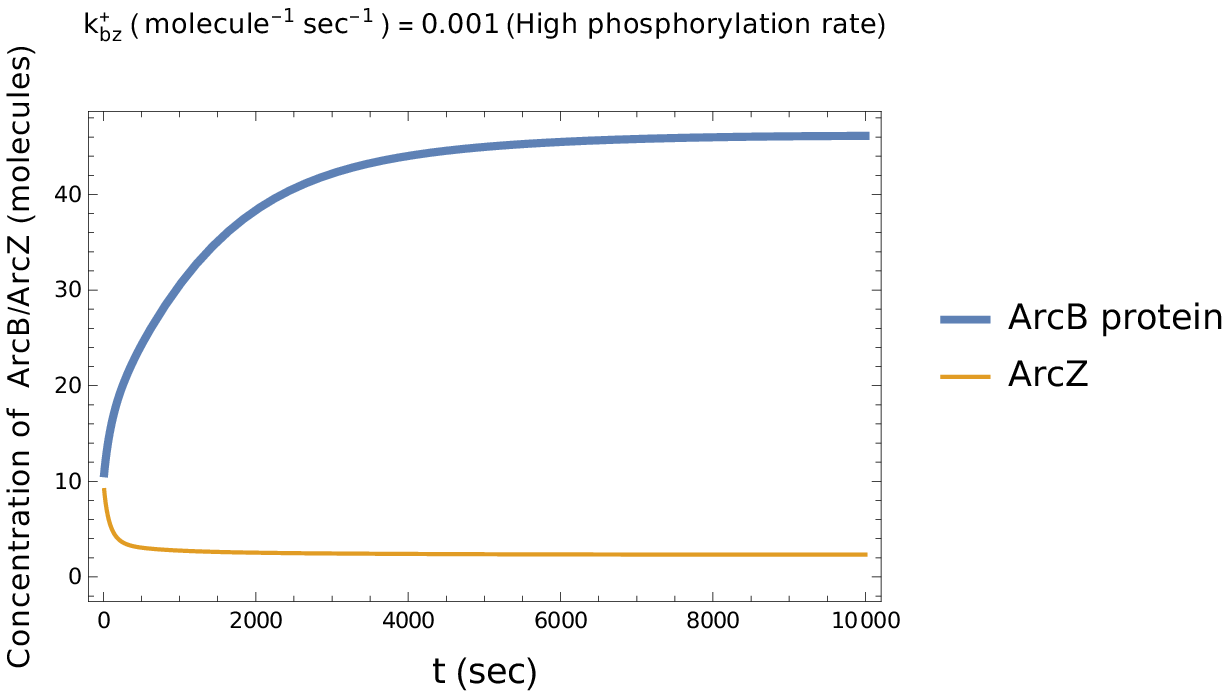}  
 \endminipage
 \caption{ Time evolution of the number of ArcB and ArcZ molecules per cell for a high phosphorylation rate with 
 $k_{\rm AP}=0.1\ {\rm molecule}^{-1}$. The remaining parameter values are as mentioned in  table \ref{table-one}. }
\label{fig:highb-z_math}
\end{figure}

\begin{figure}[!htb]
  \minipage{0.5\textwidth}
 \includegraphics[height=0.55\textwidth]{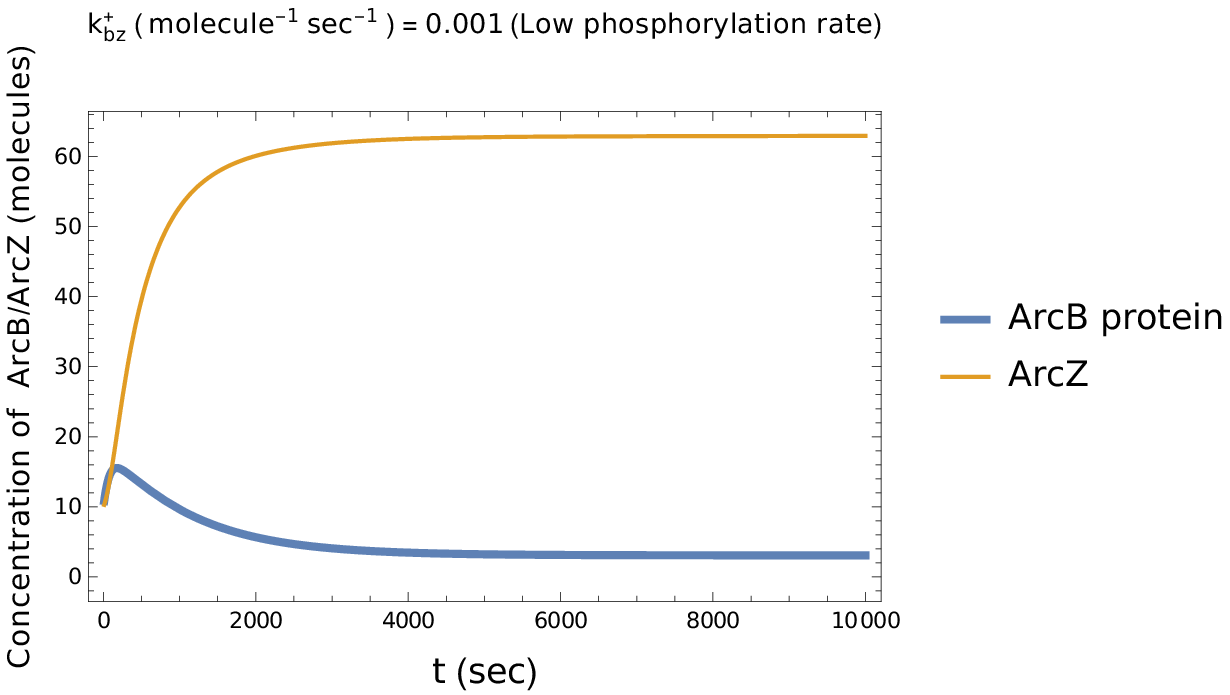}
 \endminipage
 \minipage{0.5\textwidth}
  \includegraphics[height=0.55\textwidth]{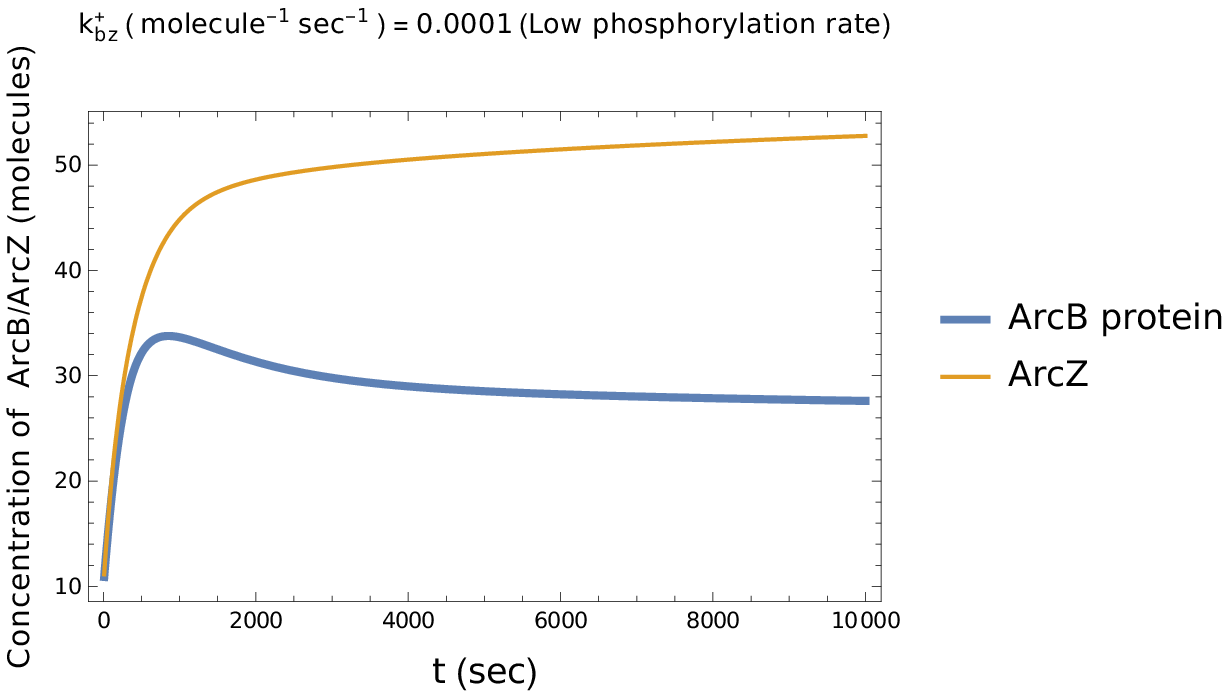}  
 \endminipage
  \caption{ Time evolution of  the number of ArcB and ArcZ  molecules per cell for a low phosphorylation rate with 
 $k_{\rm AP}=0.001\  {\rm molecule}^{-1}$. The remaining parameter values are as mentioned in  table \ref{table-one}.  }
\label{fig:lowb-z_math}
\end{figure}

In  a quasi steady-state approximation, we solve 
numerically the  time-dependent differential equations (\ref{diffeqnbm}), (\ref{diffeqnbp}) and 
(\ref{diffeqnz})-(\ref{diffeqnsigmamz}),  assuming  equilibrium condition  for the  phosphorylation kinetics. 
The parameter values used to obtain the solution are displayed in table (\ref{table-one}). These are the average 
 values  obtained from references \cite{alon, shimoni,gadgil}. sRNA synthesis 
rate is considered to be $5$ times larger than the  mRNA synthesis rates.  Values of $k_z^+$ and $k_\sigma^+$ are the 
average values obtained from \cite{alon}. 
Since the detachment rates typically have wide variations depending  on the bond strength, we 
assume $k_z^-,\ k_\sigma^-=1.5\ {\rm sec}^{-1}$ ($>1\ {\rm sec}^{-1}$) \cite{alon}.  Using the average  values for 
phosphorylation and dephosphorylation rates (see reference \cite{miller}), we find $k_{\rm AP}=0.004\ {\rm molecule}^{-1}$. 
For low and high  phosphorylation 
rates, we choose $k_{\rm AP}=0.001,\  0.1\  {\rm molecule}^{-1}$, respectively. 
The  nature of the  approach to  equilibrium    and also the equilibrium  
concentrations of various components for high and low phosphorylation rates 
are displayed in  figures  (\ref{fig:highb-z_math}),
and (\ref{fig:lowb-z_math})).

\begin{figure}[!htb]
 \includegraphics[height=0.3\textwidth]{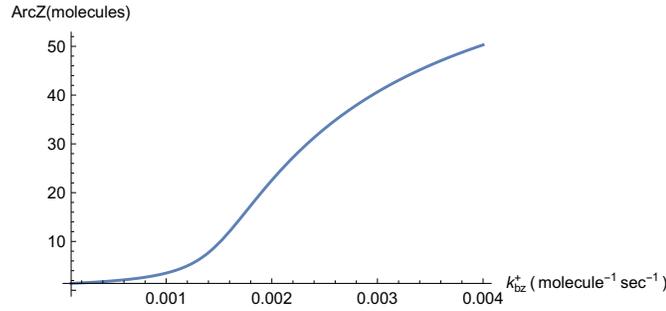}
  \caption{ Number of ArcZ molecules per cell for different values of $k_{\rm bz}^+$  for  a high phosphorylation rate, 
  $k_{AP}=0.1\  {\rm molecule^{-1}}$. The remaining parameter 
  values are as given in table \ref{table-one}. }
\label{fig:arcz-kbzp}
\end{figure}

In the high phosphorylation state, the concentration of ArcB is expected to be higher than ArcZ. However, the  differential 
equation based analysis   shows that  the approach to a high ArcB or a  high ArcZ state  
crucially depends on the  sign of  the $(\gamma_{bm}+K[A]-\frac{\beta_z}{\gamma_z} G_z^{\rm tot} k_{bz}^+)$
term  in (\ref{steadyz}). For large positive values of this term, which may hold good  for high phosphorylation rates, 
we have a low concentration of ArcZ,  
\begin{eqnarray}
[Z]\approx4 \frac{\gamma_{bm}}{\gamma_z}\beta_z G_z^{\rm tot}/(\gamma_{bm}+K[A]-
\frac{\beta_z}{\gamma_z} G_z^{\rm tot} k_{bz}^+) \label{approxz}
\end{eqnarray}
 and consequently  a high  concentration of ArcB 
due to equations (\ref{steadybm}) and (\ref{steadybp}). Further,   equation (\ref{approxz})
 shows that  for small
binding affinity between ArcZ sRNA and  ArcB mRNA  i.e. $k_{bz}^+$
 ($k_{bz}^+<<(\gamma_{bm}+K[A)/(\frac{\beta_z}{\gamma_z} G_z^{\rm tot})$), the concentration 
$[Z]$ becomes almost  independent of $k_{bz}^+$ (see figure (\ref{fig:arcz-kbzp})).
By  tuning  $k_{bz}^+$, one may, however,  see a transition to 
a high ArcZ state even when the phosphorylation rate is high for which 
 originally a high ArcB state would be  expected. 
The  steady increase 
in ArcZ is seen when $(\gamma_{bm}+K[A]-\frac{\beta_z}{\gamma_z} G_z^{\rm tot} k_{bz}^+)$  becomes negative 
with increasing $k_{bz}^+$.  
In the low phosphorylation 
case, a similar change from a high ArcZ state to a high ArcB state is seen as the value of $k_{bz}^+$ is decreased below a definite value.
The result in equation (\ref{steadyz})  and  figure (\ref{fig:arcz-kbzp})  may  be compared with the smoothening of the 
threshold-linear response  seen in case of 
sRNA mediated mRNA degradation \cite{hwa2}.  In the   threshold-linear response, the gene expression is 
completely silenced when the target transcription rate is below a threshold. Above the threshold, 
mRNAs code for the protein leading to a linear increase in the  protein 
concentration  with the target  transcription rate.  In case of strong sRNA, mRNA interaction,  
the transition from one gene expression regime  to the other  is sharp.  As the  interaction between sRNA and mRNA 
becomes weaker, the transition becomes smoother although the  threshold-linear form is preserved.  
 Figure (\ref{fig:arcz-kbzp}) describes a   similar  smooth transition from a negligible ArcZ expression regime  
 to a high ArcZ expression regime except for the fact that beyond the threshold the change in concentration with 
 $k_{bz}^+$ is not linear. 
 As we have discussed later, similar threshold behavior cannot be  found  by varying any other parameter 
 present in equation (\ref{steadyz}). Since the sRNA and mRNA binding affinity can be modulated through the 
 use of appropriate RNA chaperone,  $k_{bz}^+$ appears to be an important control parameter for achieving the 
 right concentration level  of ArcZ sRNA and  ArcB protein.

  \subsection*{Role of sRNA mediated destabilisation}
\begin{figure}[!htb]
  \minipage{0.5\textwidth}
 \includegraphics[height=0.55\textwidth]{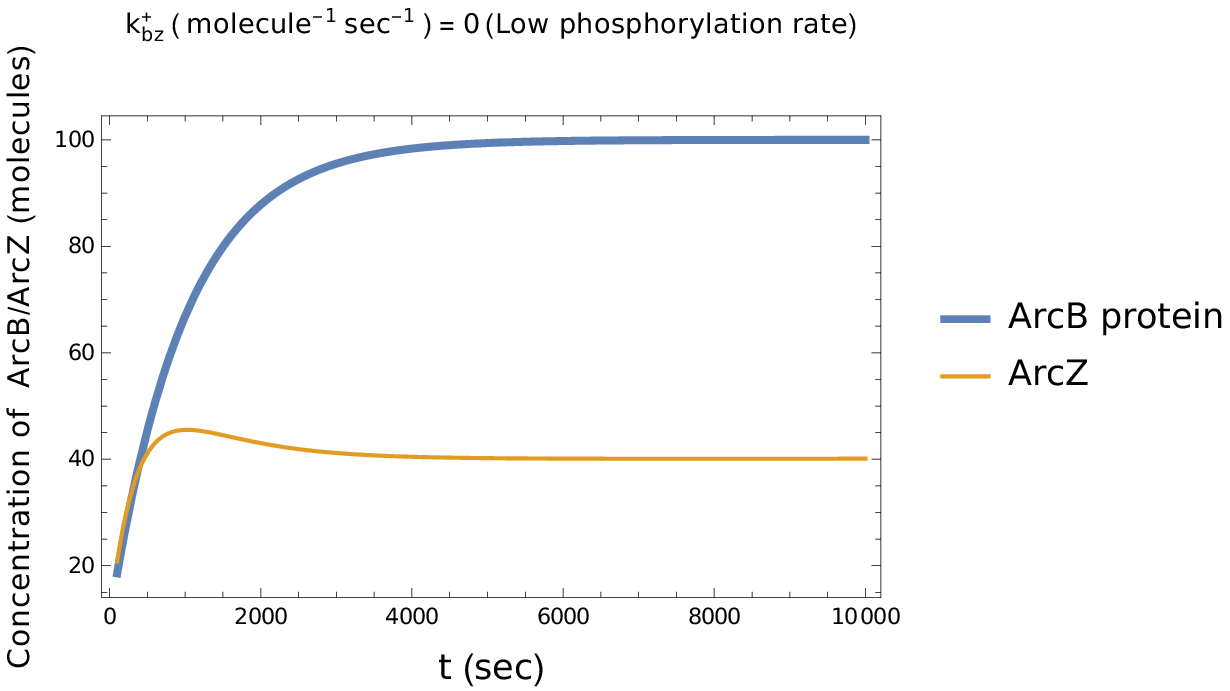}
 \endminipage
 \minipage{0.5\textwidth}
  \includegraphics[height=0.55\textwidth]{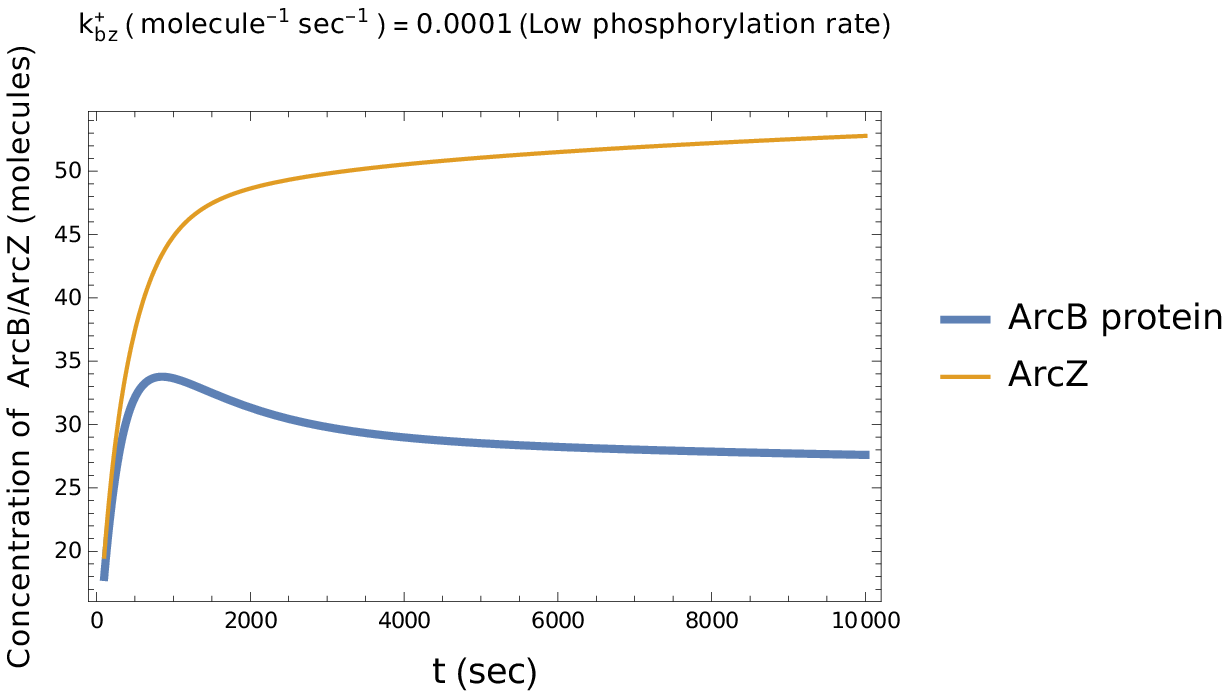}  
 \endminipage
 \caption{ The phosphorylation rate is chosen to be low, $k_ {\rm AP}=0.001 \ {\rm molecule}^{-1}$. The y-axis represents the 
 number of ArcB or ArcZ molecules per cell.  The 
 remaining parameter values are as mentioned in  table \ref{table-one}. }
\label{fig:role-of-sRNA}
\end{figure}

 In order to understand the specific role of sRNA mediated regulation, we first consider how the concentrations 
ArcZ and ArcB protein change with time in the absence of sRNA mediated mRNA destabilisation for high and low 
phosphorylation rates. In the absence of sRNA-mRNA interaction, the  equilibrium concentration of ArcB protein and ArcZ are 
 $B_p=\frac{r_{bp} r_{bm}}{\gamma_{bp}\gamma_{bm}}$ and $[Z]=\frac{\beta_z}{\gamma_z} \frac{G_z^{\rm tot}}{1+k_z [AP]}$ where 
 $[AP]$ is the concentration of the phosphorylated ArcA (see Supporting Information). 
 While for high phosphorylation rate the concentration of ArcZ reduces continuously, for low phosphorylation rate it approaches 
 a maximum saturation  value proportional to  $\beta_z/\gamma_z$.  Clearly, this saturation to the maximum value, 
 in the low phosphorylation state, arises due to the catalytic nature of 
 transcriptional repression. Thus,  whether ArcZ or ArcB has  higher concentration  
 in the low phosphorylation scenario 
 is determined  completely by the
   synthesis and degradation rates mentioned above (see figure (\ref{fig:role-of-sRNA})).  There is a  significant  increase and 
 decrease in ArcZ and ArcB concentrations, respectively, as  the 
 ArcZ mediated ArcB destabilisation is turned on (the  second panel of figure (\ref{fig:role-of-sRNA})).  
 From this observation it appears that the while, in the high phosphorylation case, the phosphorylation rate 
 alone can reduce  the concentration of  ArcZ sRNA, for low phosphorylation case an 
 sRNA mediated destabilisation of mRNA might be  important for reaching the required level of   concentrations. 
 Further in view of the fact that  the synthesis and degradation rates are genetically controlled, 
 this additional control through the   binding 
 affinity  seems to be an effective means of ArcZ/ArcB regulation for the low phosphorylation case. 

\subsection*{Approach to equilibrium}
Below, we analyse the approach to the equilibrium by linearising equations 
(\ref{diffeqnbm}), (\ref{diffeqnbp}) and (\ref{diffeqnz})-(\ref{diffeqnsigmamz}) about the equilibrium 
concentrations.   The approach to the equilibrium  can be described by the following matrix 
equation 
\begin{eqnarray}
\frac{d{\bf X}}{dt}={\bf A} {\bf X},
\end{eqnarray}
where ${\bf X}$ is a  column matrix having elements that denote small deviations from the 
equilibrium concentrations and ${\bf A}$ is a coefficient matrix with elements dependent on 
various rate constants and 
equilibrium concentrations denoted as $[..]^*$. Explicit forms of these matrices are 
{\bf X}= \[ \left( \begin{array}{c}
\delta [B_m]\\
\delta [B_p]\\
\delta [Z]\\
\delta [\sigma^s m]\\
\delta[\sigma^s mZ]\\
\delta [B_mZ]
\end{array} \right)\] 
and 
\begin{adjustwidth}{-2.25in}{0in}
{\bf A}= \[ \left( \begin{array}{cccccc}
-(k_{bz}^+[Z]^*+\gamma_{bm})& 0 & -k_{bz}^+ [B_m]^* & 0 & 0 & 0 \\
r_{bp} & -\gamma_{bp} & 0 & 0 & 0  & 0  \\
-k_{bz}^+ [Z]^* & -\frac{\beta_z G_z k_z k_{AP} [A]}{(1+k_z k_{AP} [A] [B_p]^*)^2} & -(k_{bz}^+ [B_m]^*+\gamma_z+k_{mz}^+ [\sigma^sm]^*)
 & -k_{mz}^+ [Z]^* & k_{mz}^- & k_{bz}^-\\
 0 & -\frac{\beta_\sigma G_\sigma k_\sigma k_{AP} [A]}{(1+k_\sigma k_{AP} [A] [B_p]^*)^2} & -k_{mz}^+ [\sigma^s m]^* & -(\gamma_{m}+k_{mz}^+ [Z]^*) &
 k_{mz}^- & 0 \\
 0 & 0 & k_{mz}^+ [\sigma^sm]^* &  k_{mz}^+ [Z]^* & -k_{mz}^- & 0\\
 k_{bz}^+ [Z]^* & 0 & k_{bz}^+ [B_m]^* & 0 & 0 & -k_{bz}^-
 \end{array} \right). \]
 \end{adjustwidth}
 The eigenvalues of this matrix  give an estimate of  the time scale  over which the system reaches equilibrium.  The dependence 
 of several of the  matrix  elements  on $k_{mz}^+$ and $k_{mz}^-$ (the binding  and dissociation   rates of ArcZ sRNA - $\sigma^s$ mRNA complex) 
 indicate  that the   eigenvalues  i.e. the time scale  depends on  these rate constants.  The numerical evaluation of the eigenvalues indicates 
 that the approach to the  equilibrium state is faster  as the values of $k_{mz}^+$ and $k_{mz}^-$ are decreased and increased,  
 respectively or, in other words, as the $\sigma^s$-ArcZ complex becomes more short lived.

 \subsection{Parameter sensitivity}
 The local parameter sensitivity  to different parameters can be directly obtained by  differentiating  the 
 concentration  with respect to appropriate parameters. In order to have an idea in terms of the numbers as how various 
 parameters affect the concentration of ArcZ,  we have increased or decreased  various parameter values by 
 $10\%$ about the values listed in  table \ref{table-one}.  While  the change in the concentration of ArcZ remains approximately 
 within $25\%$  for $10\%$ change in $r_{bm}$, $\gamma_{bm}$ and  $r_{bp}$, there might be a change in concentration 
 (between $25\%-35\%$ ) as the parameters related to ArcZ synthesis, degradation, concentration 
 of ArcA and degradation of ArcB protein (not mediated by ArcZ)  are changed by $10\%$.  The most significant change in ArcZ concentration 
 ($34\%$) is seen as the the ArcZ degradation rate is altered.
 
\begin{figure}[!htb]
 \includegraphics[height=0.3\textwidth]{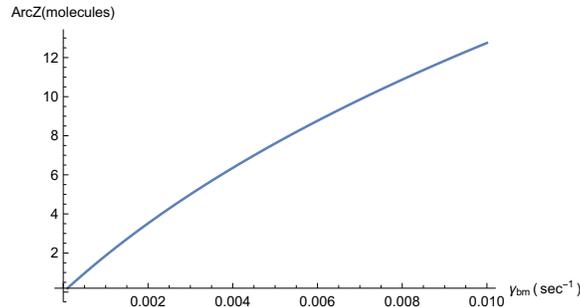}
  \caption{ Number of ArcZ molecules per cell for different values of $\gamma_{bm}$  for  a high phosphorylation rate, $k_{AP}=0.1\ {\rm molecule}^{-1}$. The remaining parameter 
  values are as given in table \ref{table-one}. }
\label{fig:arcz-gambm}
\end{figure}

 Further, it might be emphasised that overall, the most abrupt change in concentration is brought about by $k_{bz}^+$. This 
 is evident from the comparison of figure (\ref{fig:arcz-kbzp}) and  (\ref{fig:arcz-gambm}) (figure (\ref{fig:arcz-gambm}) does not have 
 distinct low and high  expression regimes). Equation (\ref{steadyz}) indicates that 
 other  parameters except $k_{bz}^+$ would also  
 give rise to similar increase or decrease   in concentration  as shown in figure (\ref{fig:arcz-gambm}) without any threshold 
behavior in particular. 
From this it appears that  in addition to the phosphorylation rate,  
 $k_{bz}^+$, i.e the strength of   sRNA and  mRNA interaction,  might be a key 
 parameter for  regulating  the  concentration levels of  ArcZ and ArcB. 
 
  \subsection*{Results from stochastic simulations}
 In the  last section, we have studied    differential  equations that describe the time evolution of the 
 average  concentrations of various proteins, RNAs etc. 
 Results from stochastic simulations using Gillespie algorithm \cite{gillespie}  describe the  noisy  nature of protein or mRNA 
 concentrations.  These results also show how the binding affinity between ArcB mRNA and ArcZ sRNA  
 influences the nature of the final equilibrium state. Since  ArcZ  sRNA upregulates its own transcription by 
 destabilising the  ArcB mRNA, one of our aims is to study  how this sRNA mediated non-catalytic reaction 
 alters the time course of  ArcZ expression whose transcription is otherwise repressed through the phosphorylation 
 activity of ArcB kinase.

 The reaction rate of the $i$th reaction listed in the Supporting Information
 is denoted below as $r(i)$. Results of stochastic simulations  done for    
parameter values  $r(1)=r(2)=r(3)=r(4)=0.001$,  $r(5)=0.001$, $r(6)=0.003$, $r(7)=0.02$, $r(8)=0.001$, 
$r(9)=0.003$, $r(10)=0.1$, $r(11)=0.8$ and $r(12)=0.02$, 
$r(13)=0.004$, $r(16)=0.002$, $r(17)=0.01$, $r(18)=0.02$, $r(19)=0.01$, $r(20)=0.001$, $r(21)=0.002$, 
$r(22)=0.001$ and $r(23)=0.0025$ are presented in the following.  Values of $r(14)$ and $r(15)$ indicating the  
association and dissociation 
rates of ArcB mRNA and ArcZ, respectively,  are changed in the simulation.

 \subsubsection*{Autoregulation by ArcZ sRNA}
In case of only transcriptionally repressed gene expression,  it is known that the 
protein synthesis is associated with transcriptional bursts which occur due to occasional 
leakage in mRNA transcription  and subsequent translation  of the mRNA leading to an abrupt increase in the protein concentration. 
In contrast to this, for sRNA mediated translational repression,   the gene expression is found to be much smoother \cite{hwa,hwa2}.
In the latter case, although transcription events are frequent, sRNAs reduce the bursts.   The difference in the noise properties are expected 
to be crucial since large transcriptional bursts  may give rise to switching  of the cell from one state to another \cite{madanbabu}. A quantitative description  
of the noise characteristics can be obtained through the probability distribution of  the protein concentrations over a time interval.  
It  has been found that the probability distributions  in case of sRNA mediated translational repression has less variance in comparison 
with the transcriptional repression \cite{hwa2}.  
     
\begin{figure}[ht!]
\centering
 \includegraphics[height=0.4\textwidth]{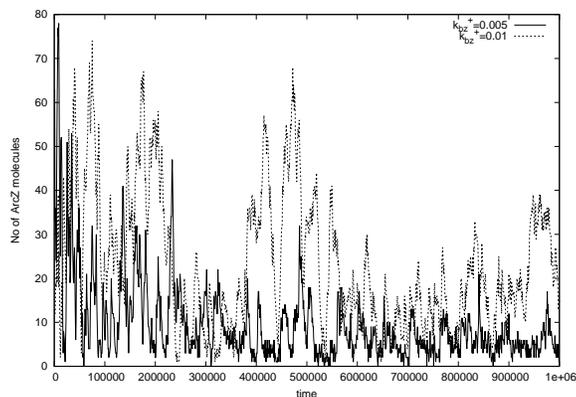}
\caption{ ArcZ synthesis pattern for two different values of $k_{bz}^+$ in case of low phosphorylation rate with $r(1)=r(2)=0.001$.}
\label{fig:kbzdependence}
 \end{figure}

\begin{figure}[ht!]
\centering
 \includegraphics[height=0.4\textwidth]{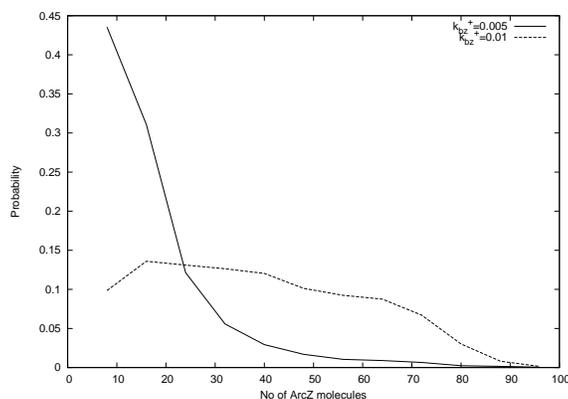}
\caption{Probability distributions of ArcZ concentration obtained from  stochastic simulation of the  reaction processes for two different values of $k_{bz}^+$. 
The distributions are obtained upon averaging over $20$ cells (samples).  
A low phosphorylation rate  with $r(1)=r(2)=0.001$ is considered here.  }
\label{fig:probdist}
 \end{figure}

 In order to understand  how the autoregulation by ArcZ sRNA influences the temporal behavior of its own synthesis, 
 we show  how   the time course of  {\it arcz}  expression  during a single run of stochastic simulation  changes as 
 the  ArcZ-ArcB binding affinity is changed (see figure (\ref{fig:kbzdependence})).   The figure shows that  with the 
 increase in $k_{bz}^+$, transcriptional bursts with widely varying strengths become more frequent. 
 In order to further understand the noise characterisitcs  for different   values of $k_{bz}^+$, we have 
 obtained  the probability distributions of  the number of ArcZ molecules over  the time course of ArcZ synthesis during a  fixed time interval. 
 Probability distributions obtained upon averaging over  20 cells (samples) are displayed in figure (\ref{fig:probdist}). 
 With the increase in ArcZ mediated ArcB destabilisation,  the probability distribution becomes flatter due to frequent transcriptional 
 bursts of varying strengths (see figure (\ref{fig:kbzdependence})).

\begin{table}[ht]
\begin{adjustwidth}{-2.25in}{0in}
\caption{Parameter Values}
\centering
\begin{tabular} {|c| c| c| }
\hline\hline
Reaction & Rate Constants & Parameter Values\\
\hline
ArcZ synthesis  & $\beta_z (*)$  & 0.1\ ({\rm molecule/sec})\\
\hline
ArcZ degradation & $\gamma_z$  &0.0025 \ (${\rm sec^{-1}}$) \\
\hline
$\sigma^s$-mRNA synthesis & $\beta_\sigma$ & 0.02 \ ({\rm molecule/sec}) \\
\hline
$\sigma^s$-mRNA degradation & $\gamma_m$  & 0.002\ (${\rm sec^{-1}}$)\\
\hline
ArcB mRNA synthesis & $r_{bm}$  &  0.02 \ ({\rm molecule/sec})\\
\hline
ArcB mRNA degradation &  $\gamma_{bm}$   & 0.002\ (${\rm sec^{-1}}$)\\
\hline
ArcB protein synthesis & $r_{bp}$ & 0.01 \ (${\rm sec^{-1}}$)\\
\hline
ArcB protein degradation & $\gamma_{bp}$ & 0.001\ (${\rm sec^{-1}}$) \\
\hline
ArcB-ArcZ mRNA complex formation & $k_{bz}^+$  & 0.01\  (${\rm molecule^{-1} sec^{-1}}$) \\
\hline
ArcB-ArcZ mRNA complex dissociation (ArcZ mediated degradation of ArcB) & $k_{bz}^-$  & 0.01\ (${\rm sec^{-1}}$) \\
\hline
$\sigma^s$mRNA-ArcZ complex formation   &  $k_{mz}^+$ & 1 \  (${\rm molecule^{-1} sec^{-1}}$) \\
\hline
$\sigma^s$mRNA-ArcZ complex dissociation  & $k_{mz}^-$  & 0.02 \ (${\rm sec^{-1}}$)\\
\hline
Repression of arcZ gene by ArcA-P & $k_z=k_z^+/k_z^-(*)$  & 0.1\ (${\rm molecule^{-1} }$)\\
\hline
Repression of $\sigma^s$ gene by ArcA-P & $k_\sigma=k_\sigma^+/k_\sigma^-(*)$   & 0.1 \ (${\rm molecule^{-1} }$)\\
\hline
Phosphorylation activity & $k_{AP}=\frac{k_p k_c}{1+k_p^+/k_c^-} (*)$ &0.1  {\rm or} 0.001 \ (${\rm molecule^{-1} }$)\\
\hline
\end{tabular}\label{table-one}\\
We have chosen an equilibrium concentration for ArcA molecules, $[A]=60 \ {\rm molecules}$.  The parameter values are obtained 
from  references \cite{shimoni} and \cite{alon}.  For details  on parameters with $*$, see the discussion in the main text. 
\end{adjustwidth}
\end{table}
\section*{Discussion} 
\label{sec:discussion}
In this paper, we have studied how transcriptional and translational regulation combinedly 
influence  protein and sRNA synthesis in a particular stress response network of {\it E.coli}. 
This network   regulates the synthesis of  the $\sigma$-factor, $\sigma^s$, in response to oxygen 
and energy availability to the cell.  
It has been found experimentally  that 
this regulatory network  functions through 
feedforward as well as feedback mechanisms involving ArcA, ArcB proteins and ArcZ sRNA as 
the major regulators. The phosphorylation activity of   ArcB kinase depends on the oxygen and energy 
availability to the cell.  Upon phosphorylation by ArcB, the phosphorylated ArcA protein functions  
as a transcriptional repressor for $\sigma^s$ as well as for ArcZ which activates $\sigma^s$ synthesis post transcriptionally. 
The regulation of $\sigma^s$ synthesis by ArcA thus happens through a feedforward mechanism. In addition to this, 
ArcZ sRNA  destabilises ArcB mRNA  through a feedback mechanism.   
In case of aerobic growth, the phosphorylation activity 
of ArcB is reduced. As a consequence of this, ArcZ  synthesis is enhanced causing  enhanced downregulation 
of ArcB. In case of anaerobic growth, the rate of ArcA phosphorylation increases causing a 
 significant repression of ArcZ synthesis and hence  a high ArcB concentration. 
 
 The main purpose of the present study is to explore   the combined 
 effect of catalytic and non-catalytic interactions associated with 
  transcriptional and translational regulation, respectively. 
 Some of the recent studies \cite{hwa,hwa2} 
 have demonstrated that sRNA mediated non-catalytic regulation may 
 give rise to interesting features such as tunable threshold-linear response of gene expression, smoothening of  fluctuations in gene expression   etc. 
 In view of these results, it seems interesting to ask what the combined effect of both transcriptional and translational regulation 
 would be on various  gene expressions. This analysis  also helps us  understand  the role of sRNA in attaining certain properties 
 which cannot be  attained otherwise through transcriptional regulation alone.

Our main observations are listed in the following.
Qualitatively, depending on the phosphorylation activity, the system can   have a   high 
ArcZ (consequently a low ArcB) state or a low ArcZ (high ArcB) state.  (a) Based on  experimental  observations, 
earlier, there were speculations that this regulation mechanism could give rise to bi-stability. 
The present analysis, however,  leads to  a single equilibrium solution for ArcZ concentration which varies 
continuously from a high to a low  value as, for example, the phosphorylation rate is increased. ArcB concentration, 
being inversely related to ArcZ concentration, correspondingly increases as the phosphorylation rate is increased.   
In addition to the phosphorylation rates, the equilibrium  concentrations  depend   on 
several other parameters such as various synthesis  and degradation rates and the binding affinity between the ArcZ sRNA 
and its target, ArcB mRNA.   We show that, among various parameters, the binding affinity 
between the sRNA and mRNA plays a distinct role  in regulating the concentrations. In particular,  
in the high phosphorylation state,  the dependence of 
ArcZ concentration on the binding affinity is similar to the smooth threshold-linear mode found earlier \cite{hwa,hwa2}. 
 In the present case,  however, the dependence of  ArcZ concentration on the binding affinity beyond the threshold 
 is not linear.   For example, although in 
 case of high phosphorylation rate one expects
 a low ArcZ  concentration, this might not be possible in case of  strong destabilisation of ArcB mRNA by ArcZ sRNA.  
 A strong destabilisation of ArcB by ArcZ leads to low ArcA/ArcB  mediated repression of ArcZ synthesis resulting in a high ArcZ 
 concentration.  In a similar manner, in case of a low phosphorylation rate, a high ArcZ state may not be achievable if  ArcZ mediated 
 destabilisation of ArcB mRNA is weak.  As has been shown, this kind of threshold behavior  can affect the concentration levels abruptly. 
 Although other synthesis and degradation rates influence the concentration levels significantly, 
 none of these  rate constants shows such threshold behavior. 
  Further, given that the synthesis and degradation rates of  various proteins and mRNAs are genetically controlled, this kind of 
 threshold behavior  due to sRNA-mRNA interaction might be particularly beneficial for stress response.
  (b) In contrary to the above observation,  the binding affinity between ArcZ sRNA and $\sigma^s$ mRNA  has marginal influence on 
 the equilibrium state. We show that this interaction influences only the time scale over which the system approaches the equilibrium. 
 (c) We show that the sRNA mediated destabilisation of mRNA might be particularly crucial in the low phosphorylation state. 
 Since ArcZ synthesis is transcriptionally repressed by phosphorylated ArcA, due to the  catalytic nature of transcriptional 
 repression, ArcZ concentration saturates to an upper limit in the low phosphorylation state.  
 We show that the sRNA mediated  target destabilization is crucial here for  further increase in ArcZ  concentration. 
 The  sRNA-mRNA interaction might not be as crucial  as this in case of high phosphorylation 
 state   since, in this state, the  transcriptional repression alone is enough to reduce  ArcZ concentration to any small value. 
 (d) In order to understand how ArcZ sRNA autoregulates its synthesis by destabilising  ArcB mRNA, we find  
  the probability distributions for ArcZ concentration from  the noisy gene expression data obtained from
   stochastic simulations over a given time interval. The average probability distribution (averaged over $20$ samples) 
   are obtained  for different values of ArcB-ArcZ binding affinities. 
   With the increase in  ArcB-ArcZ interaction,  the probability distribution becomes flatter  indicating increased number of 
   transcriptional bursts with varying strengths. 
 This feature might have larger implications in terms of switching of the cell from one stable state to another \cite{madanbabu}.   
  
  Since the focus of the  present work is on the role of ArcZ sRNA  on regulating  the concentration levels 
 of ArcB and ArcZ, we have not considered a third layer  of regulation of $\sigma^s$ through
 the proteolysis of $\sigma^s$.  
  It has been found that, in a branched pathway,  ArcB phosphorylates ArcA as well as another 
  response regulator RssB  which is a proteolytic target factor for $\sigma^s$  \cite{mika}. 
 Although ArcA phosphorylation rate is found  to be  $10$ times faster than  that of RssB,  it is 
 believed that they compete for phosphotransfer from ArcB and RssB phosphorylation might be 
 sensitive to slight changes in  ArcA concentration in case of a limited supply  of ArcB-P.   
 Further, it is also found that  $\sigma^s$ might affect the expression level of 
 ArcA \cite{mika}. In the present analysis,  we have restricted 
 ourselves to $\sigma^s$ transcripts and we have 
 not considered the translation part. At this level, this branched pathway and $\sigma^s$ mediated 
 regulation of ArcA expression may affect  our main result on ArcZ concentration
 by altering effectively the ArcA and ArcB concentration by a fraction without changing the main conclusion on 
 the role of ArcZ mediated regulation.  
 Since, the regulation by  RssB and the associated competition for phosphotransfer can  
 lead to  an interesting regulation mechanism  for $\sigma^s$, an extended analysis of $\sigma^s$ regulation 
 will be published elsewhere. Additionally, it would be interesting to  combine this along with a detailed stochastic analysis to 
find  what the impact of wide stochastic fluctuations in ArcZ synthesis would be on $\sigma^s$ synthesis. 
 
In conclusion, the present analysis suggests  multiple roles  of sRNA in  regulating its own concentration as well as that of other 
 proteins and mRNAs.    Regulation by  sRNAs is typically due to  their  base-pairing with   mRNAs.  
 Previous studies show that this base-pairing of bacterial sRNA is  often facilitated  
 by the action of RNA chaperone, Hfq.  For example, Hfq might give rise to conformational changes in the target mRNA 
 and promote accessibility for the sRNA  and hence the binding between the mRNA and the sRNA \cite{gottesman,papenfort}. 
 Independent experimental studies have  previously revealed  ArcZ sRNA as  one of the targets of Hfq \cite{wassarman,zhang,sittka,sittka1}. 
 In Salmonella enterica, ArcZ is involved in Hfq-dependent post-transcriptional repression of tpx and sdaCB genes \cite{papenfort1}. 
The fact that the binding affinity of ArcZ sRNA and ArcB mRNA plays 
 an important role in achieving   the desired equilibrium  concentrations  indicates the need for molecules such as Hfq  or similar 
 ones to appropriately tune  the  interaction between ArcZ sRNA and ArcB mRNA. 
Additional experimental studies would help explore such possibilities.


\section*{Acknowledgement}

  I thank Abhishek Acharya for many useful discussions.

\section*{Supporting Information}

\paragraph*{S1 Appendix.}
\label{S1_Appendix}
{\bf  Reaction processes and details on the phosphorylation kinetics and the repressor activity} The appendix lists all the reactions and 
provides  differential equations describing   phosphorylation kinetics and the repressor activity of ArcA.

\end{document}